\documentclass{aa}  

\usepackage{graphicx, placeins}
\usepackage{txfonts}
\begin{document}


   \title{\ion{Mg}{ii}~h\&k fine structure prominence modelling\\ and the consequences for observations}
   \author{A. W. Peat\inst{1,2}
          \and
          N. Labrosse\inst{2}
          \and
          P. Gouttebroze\inst{3}\fnmsep\thanks{retired}
          }

   \institute{University of Wroc\l{}aw, Centre of Scientific Excellence -- Solar and Stellar Activity, Kopernika 11, 51-622 Wroc\l{}aw, Poland 
   \and 
   SUPA School of Physics and Astronomy, University of Glasgow, Glasgow, G12 8QQ, UK
   \and
   Université Paris-Saclay, CNRS, Institut d'Astrophysique Spatiale, 91405, Orsay, France
   \\
   \email{aaron.peat@uwr.edu.pl}
             }

   \date{}

 
  \abstract
   {}
   {Using 2D \ion{Mg}{ii}~h\&k solar prominence modelling,  our aim is to understand the formation of complex line profiles and how these are seen by the Interface Region Imaging Spectrograph (IRIS). Additionally, we see how the properties of these simulated observations are interpreted through the use of traditional 1D prominence modelling.}
   {We used a cylindrical non-local thermodynamic equilibrium (NLTE) 2D complete redistribution {(CRD)} code to generate a set of cylindrical prominence strands, which we stacked behind each other to produce complex line profiles. Then, with the use of the point spread functions (PSFs) of IRIS, we were able to predict how IRIS would observe these line profiles. We then used the 1D NLTE code PROM in combination with the {Cross Root Mean Square} method (xRMS) to find the properties recovered by traditional 1D prominence modelling.}
   {Velocities of magnitude lower than 10~km~s$^{-1}$ are sufficient to produce asymmetries in the \ion{Mg}{ii}~h\&k lines. However, convolution of these with the PSFs of IRIS obscures this detail and returns standard looking single peaks. By increasing the velocities by a factor of three, we recover asymmetric profiles even after this convolution. The properties recovered by xRMS appear adequate at first, but the line profiles chosen to fit these profiles do not satisfactorily represent the line profiles. This is likely due to the large line width of the simulated profiles.}
   {Asymmetries can be introduced by multithread models with independent Doppler velocities. The large line width created by these models makes it difficult for traditional 1D forward modelling to find good matches. This may also {demonstrate} degeneracies in the solution recovered by {single-species} 1D modelling.}


   \maketitle

\section{Introduction}
The modelling of solar prominences is crucial to our understanding of their plasma properties and the wider solar atmosphere. Without modelling we cannot fully understand what is happening inside a prominence from observation alone. One of the open questions in prominence physics put forward by \cite{schmieder_open_2014} is how complex line profiles are produced in solar prominences. One suggestion is the existence of unresolvable fine structure along the line of sight. {This was explored in \cite{gunar_large_2022}, where the authors investigated the impact of both unresolvable fine structure and the incident radiation.} In this paper we wish to {further} explore this idea and demonstrate the consequences that this has on observations and how we currently interpret prominences using 1D models. To do this we use the 2D non-local thermodynamic equilibrium (NLTE) cylindrical radiative transfer (RT) code developed over a series of seven papers.

The code was first introduced by \cite{gouttebroze_radiative_2004} with a simple 1D model of a vertical cylinder suspended in the solar atmosphere. The use of accelerated lambda iteration (ALI) methods to produce accurate and efficient numerical simulations were investigated.  \cite{gouttebroze_radiative_2005} generalised the previously developed method to two dimensions -- adding azimuthal dependence with radiation from a spherical source representing the Sun. \cite{gouttebroze_radiative_2006} introduced a ten-level (plus continuum) hydrogen atom, replacing the basic two-level atom used in the two previous   papers developing the geometry and method. \cite{gouttebroze_radiative_2007} introduced time dependencies and thermal equilibrium such that these effects on solar prominences could be investigated. {\cite{gouttebroze_radiative_2008} introduced 3D velocity fields which allow us to investigate Doppler shift and Doppler brightening/dimming effects.} Three velocity fields were introduced, translational, rotational, and expanding, that  allow the user to explore how these fields can affect the line profiles produced in a prominence. \cite{gouttebroze_radiative_2009} introduced a helium and hydrogen system. The hydrogen atom here is a five-level (plus continuum) atom and the new helium atom consists of three ionisation stages with 29 levels for \ion{He}{i}, 4 for \ion{He}{ii}, and 1 for \ion{He}{iii}. This helium atom is identical to the helium atom introduced in \cite{labrosse_formation_2001}. \cite{labrosse_radiative_2016} investigated the use of multithread models to explore the formation of the {principal} Lyman, Balmer, and helium lines in solar prominences. 

In this paper we wish to further this work with a focus on the formation of the \ion{Mg}{ii}~h\&k lines in solar prominences. These lines are frequently observed by the Interface Region Imaging Spectrograph \citep[IRIS;][]{depontieu_interface_2014}, and thus employing   the point spread functions (PSFs) of IRIS allows us to explore how these fine structure (or multithread) observations are seen by the instrument,  as was done by \cite{gunar_lyman-line_2008} for the Solar Ultraviolet Measurements of Emitted Radiation \citep[SUMER;][]{wilhelm_sumer_1995} instrument on board the Solar and Heliospheric Observatory \citep[SOHO;][]{domingo_soho_1995}.  

To finish, we then use the {Cross Root Mean Square} procedure \citep[xRMS;][Peat et al. in prep.]{peat_solar_2021} to explore the parameters recovered by 1D forward modelling. This is done through the use of the 1D non-local thermodynamic equilibrium (NLTE) radiative transfer code PROM \citep{gouttebroze_hydrogen_1993, heinzel_theoretical_1994}. PROM models a static monolithic 1D slab suspended above the solar surface, and is frequently used to obtain the plasma parameters of prominence observations \citep[e.g.][]{heinzel_2008, heinzel_understanding_2015, ruan_diagnostics_2019, zhang_launch_2019, peat_solar_2021}. Here we use the version described in \cite{levens_modelling_2019}, which models the \ion{Mg}{ii}~h\&k and the \ion{Mg}{ii} triplet lines in partial redistribution (PRD).

\begin{table}
\centering
\begin{tabular}{ccc} \hline\hline
Parameter & Unit                                & Value        \\ \hline
$\alpha$  & rad                                  & $\pi/2$      \\
$r_0$     & km                                  & 500          \\
$r_1$     & km                                  & 1000         \\
$T_0$     & K                                   & 6000         \\
$T_1$     & K                                   & 100~000 \\
$P$         & dyn$~$cm$^{-2}$                     & {0.1} \\
$H$ & km & 10~000\\
$v_T$     & km~s$^{-1}$& 5  \\    \hline \hline
\end{tabular}
\caption{Parameters of the p4 model from \cite{labrosse_radiative_2016}. $\alpha$ is the rotation of the cylinder; r$_0$ is the radius of the cylinder, which is treated as isothermal and isobaric; r$_1$ is the radius (including r$_0$), which encompasses the PCTR; T$_0$ is the central temperature of the prominence; T$_1$ is the temperature of the prominence at the edge of the PCTR; P is the gas pressure; and v$_T$ is the microturbulent velocity.} 
\label{p4}
\end{table}
\section{Modelling}
In this section we   briefly cover the model used in this investigation. The code allows us to simulate a {semi-infinite} cylindrical prominence suspended above the solar atmosphere with a prominence-to-corona transition region (PCTR). However, this PCTR is only in temperature, not pressure. Pressure is instead a constant. As temperature gently rises, the static pressure causes the density to drop. The temperature profile is dictated by \citep{gouttebroze_radiative_2006}

\begin{equation}
    \log_{10} T(r)=\begin{cases}
        \log_{10} T_0, & \text{if } r \leq r_0\\
        \log_{10} T_0+(\log_{10} T_1-\log_{10} T_0)\frac{r-r_0}{r_1-r_0}, & \text{otherwise},
    \end{cases}
\end{equation}
where $r$ is the radial position in the cylinder, $r_0$ is the isothermal radius, $r_1$ is the radius of the edge of the PCTR, $T_0$ is the isothermal temperature, and $T_1$ is the temperature at the edge of the PCTR. {This essentially describes a temperature gradient across the field lines, which contradicts the field aligned PCTR described in studies such as \cite{heinzel_prominence_2001}. However, due to the semi-infinite nature of the cylinder, this is the only geometry in which the PCTR can be constructed.} The magnesium atom implemented has seven levels, one for \ion{Mg}{i}, five for \ion{Mg}{ii}, and one for \ion{Mg}{iii}. {This atom has} five {allowed} radiative transitions{; the} \ion{Mg}{ii}~h\&k lines and the three subordinate lines. This atom is effectively {identical to} that used by \cite{levens_modelling_2019}.  In this paper we   present the results pertaining to the \ion{Mg}{ii}~h line;   the analogous \ion{Mg}{ii}~k results are in  the Appendix.

\begin{figure}
    \centering
    \includegraphics[width=\linewidth]{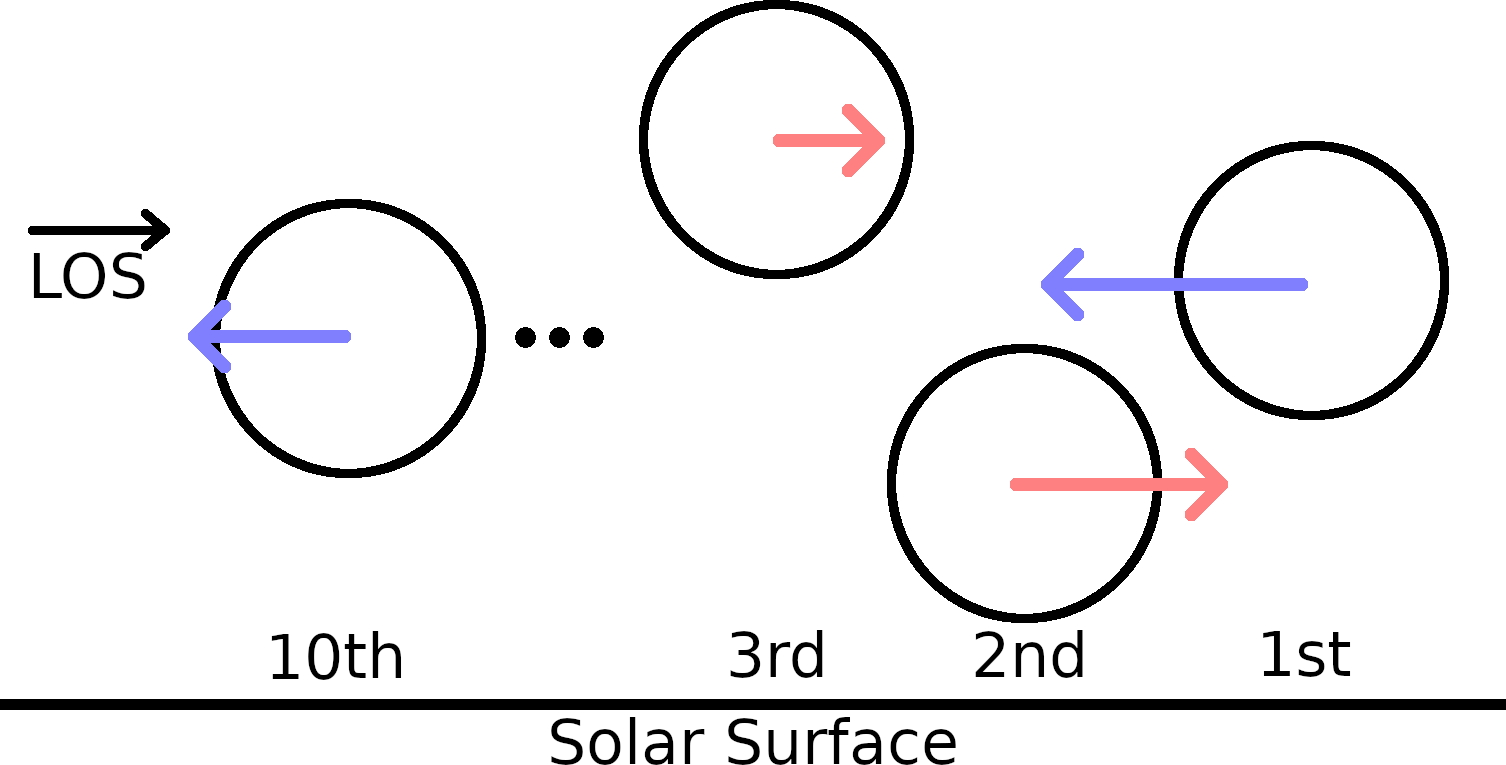}
    \caption{Configuration of the ten threads. The ten threads are shown at random displacements with random Doppler velocities. The threads are numbered from the back to the front. This figure has been adapted from \cite{peat2023PhD}.}
    \label{setup}
\end{figure}
\begin{figure}[h]
    \centering
    \includegraphics[width=0.95\linewidth]{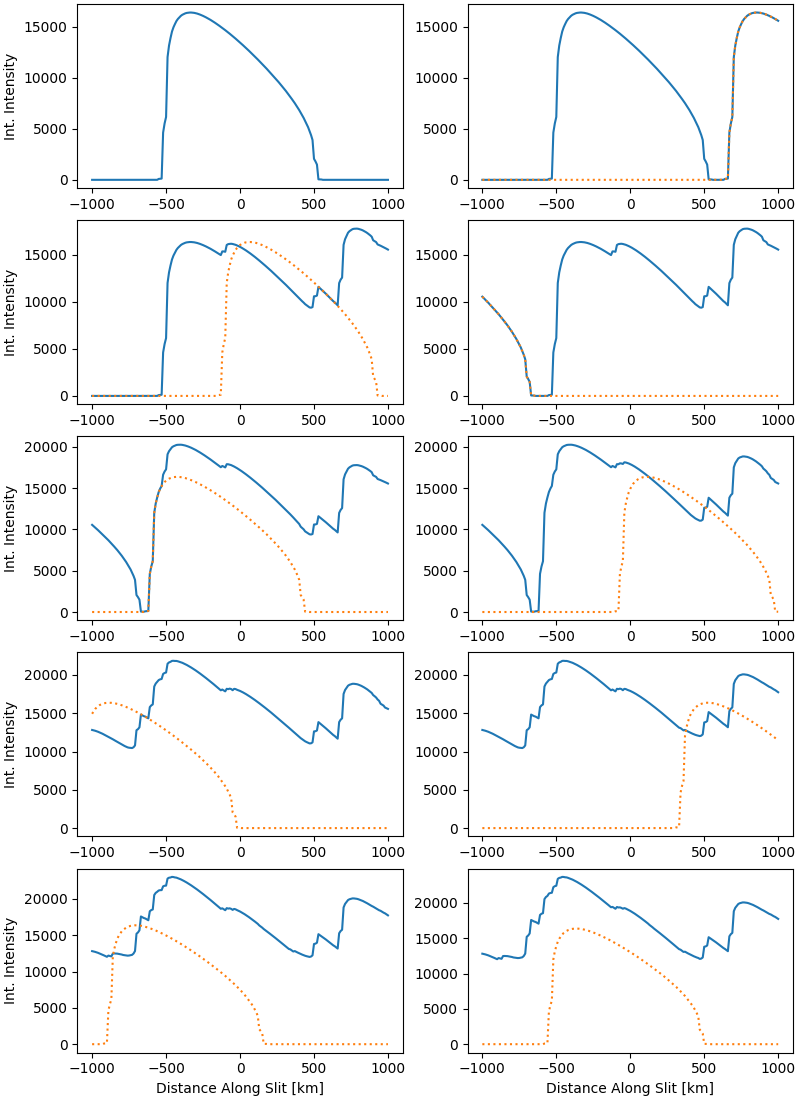}
    \caption{Effect of ten offset threads on integrated intensity along the slit. The first panel (top left)  is one thread and each subsequent panel includes one additional thread behind the first. The blue line is the {total} integrated {intensity along the slit}, and the dotted orange is the integrated {intensity} added in this panel. {The x-axis is measured relative to the centre of the front-most (10th) thread}. The units on the y-axis are erg~s$^{-1}$~cm$^{-2}$~sr$^{-1}$. This plot originally appeared in \cite{peat2023PhD}.}
    \label{fig:slith}
\end{figure}
The parameters used in this study are of the p4 model of {\cite{gouttebroze_radiative_2009} and} \cite{labrosse_radiative_2016}. Table \ref{p4} shows the parameters of this model. The inclination of the cylinder is described by $\alpha$, and is defined as the angle between the axis of rotational symmetry of the cylinder and the normal to the solar surface. This rotation is undertaken in the plane of sky. The p4 model has an $\alpha$ value of $\pi/2$, resulting in horizontally aligned threads. Multithread simulations \citep[e.g.][]{gunar_2023} tend to focus on vertically aligned threads;  this is motivated by observation. However, horizontally aligned threads are also commonly observed \citep[e.g.][]{schmieder_open_2014, levens_solar_2015,vial_observed_2016,ruan_dynamic_2018,zhang_launch_2019}, and magnetohydrodynamic simulations tend to lead to the formation of horizontal structures \citep[e.g.][]{gunar_non_2013, jenkins_resolving_2022}.

Using this p4 model, we {randomly} stack  ten threads along the line of sight {with the option of adding random line-of-sight velocities} (see  Fig. \ref{setup}). Relative to the tenth thread, which itself is at an altitude of 10,000~km, {the threads are vertically displaced by} 30, 370, -860, 550, -450, 90, 1200, -400, -1190, and 0{~km} from back to front. Each thread is simulated separately, and then the radiation from each cylinder is recursively traced through the proceeding cylinders. Mutual radiative interactions are not accounted for here, only absorption of radiation from the preceding threads. This radiation is propagated to yield the total emergent intensity, $I$, using the   equation

\begin{equation}
    I\left(\lambda\right)=I_{1}\left(\lambda\right)+I_{0}\left(\lambda\right)\exp\left(-\tau_{1}\left(\lambda, s\right)\right),
    \label{singlert}
\end{equation}
{where $I_{1}$ and $\tau_{1}$ are respectively the intensity from and the optical thickness of the front-most thread, which the radiation from the preceding thread, $I_0$, is propagating through. This can then be}
generalised for ten threads,
\begin{equation}
    I\left(\lambda\right)=I_{10}\left(\lambda\right)+\sum_{i=1}^{9}\left(I_i\left(\lambda\right)\exp\left(\sum_{j=i+1}^{10}-\tau_j\left(\lambda, s\right)\right)\right),
    \label{totalrt}
\end{equation} 
where $I_i$ is the radiation from thread $i$, and $\tau_i$ is the optical {thickness} of thread $i$ {corresponding to a path of} $s$, the path length of a ray through cylinder $i$.

\begin{figure}
    \centering
    \includegraphics[width=\linewidth]{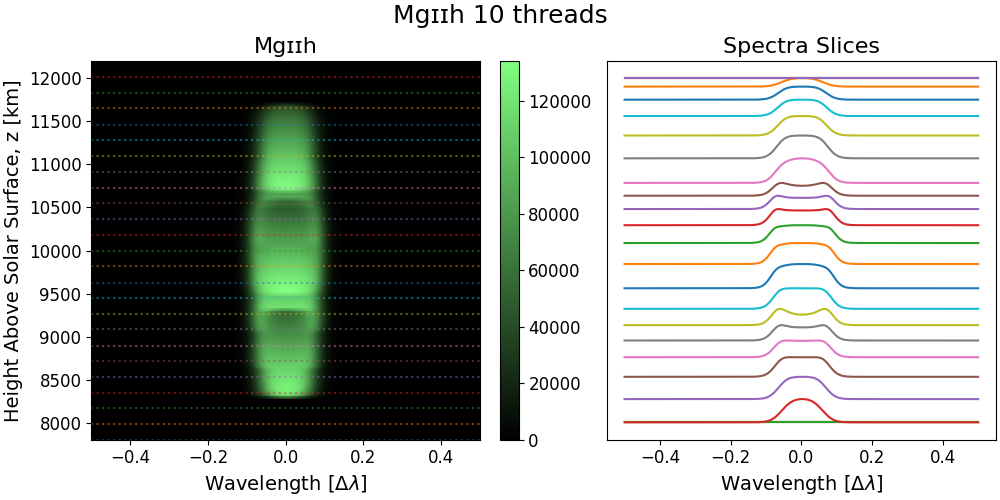}
    \caption{Two-dimensional \ion{Mg}{ii}~h spectra. Shown are the spectra of the ten stacked offset threads with example {spectral cuts}. The coloured dotted lines in the left panel correspond to the coloured solid lines on the right. This field of view of the plot has been extended to include all of the threads, but Fig. \ref{fig:slith} is restricted to the  field of view from 9000 to 11000k~m. The units of the colour bar are erg~s$^{-1}$~cm$^{-2}$~\AA$^{-1}$~sr$^{-1}$. This figure originally appeared in \cite{peat2023PhD}.}
    \label{fig:spectrah}
\end{figure}
\subsection{Static models}
\label{static}
Following the work done by \cite{labrosse_radiative_2016}, we first investigate the wavelength-integrated intensities along the slit. {Our results can be seen in Fig. \ref{fig:slith}. Compared to the figure in the previous study, we find a quite different result.} However, this study deals with \ion{Mg}{ii}~h\&k and the previous study with Ly~$\alpha$. The Ly~$\alpha$ line has considerably greater optical thickness compared to \ion{Mg}{ii}~h\&k, up to three orders of magnitude greater. The lower optical thickness of \ion{Mg}{ii}~h\&k allows more radiation from preceding threads to contribute  to  the observed spectra. The optical thickness of \ion{Mg}{ii}~h\&k is {not negligible}, however, and so most  of the radiation seen comes from the front-most threads. Figure \ref{fig:spectrah} shows the full 2D spectra of the slit with samples of the {spectra} at evenly spaced intervals of 250km ($\sim1/3$\arcsec at 1AU). While there is evidence of multithread interactions in this figure, it is not clear how many there are from either panel of this figure. The integrated intensity along the slit gives a better indication of the number of threads in the observation, but the number remains unclear. It provides a better proxy than the line profiles alone, however, which could be mistakenly identified as fewer threads due to their relative incomplexity. 
\begin{figure}
    \centering
    \includegraphics[width=\linewidth]{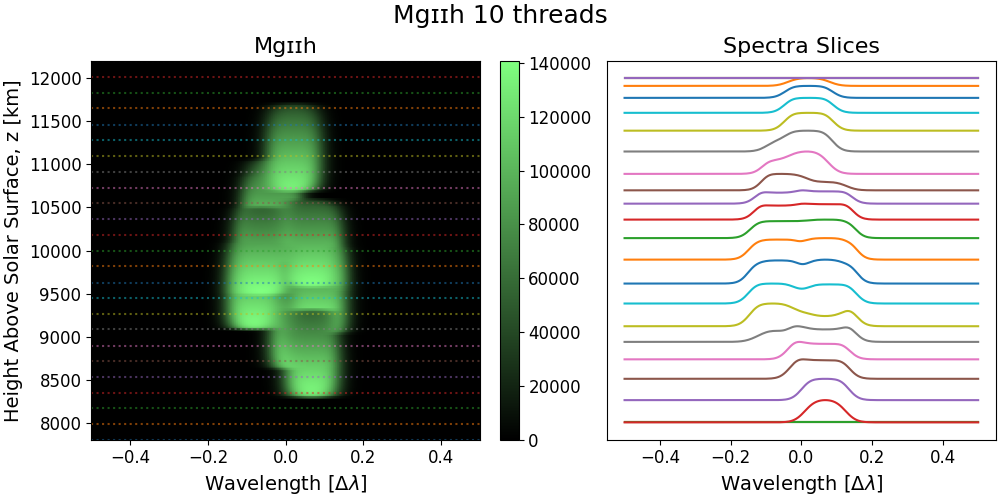}
    \caption{Similar to Fig. \ref{fig:spectrah}, but the threads have random Doppler velocities applied. The units of the colour bar are erg~s$^{-1}$~cm$^{-2}$~\AA$^{-1}$~sr$^{-1}$. }
    \label{fig:spectrahv}
\end{figure}

\begin{figure}[h]
    \centering
    \includegraphics[width=\linewidth]{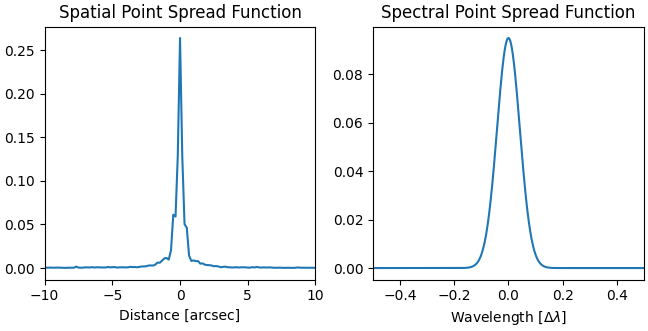}
    \caption{Point spread functions   of IRIS. Left: Spatial PSF (along the slit); Right: Spectral PSF. {The spatial PSF function has a greater extent than that shown in the plot; the x-axis is scaled to focus on the core of the function.}}
    \label{fig:psfs}
\end{figure}

\begin{figure*}
    \centering
    \resizebox{\hsize}{!}
    {\includegraphics[width=\linewidth]{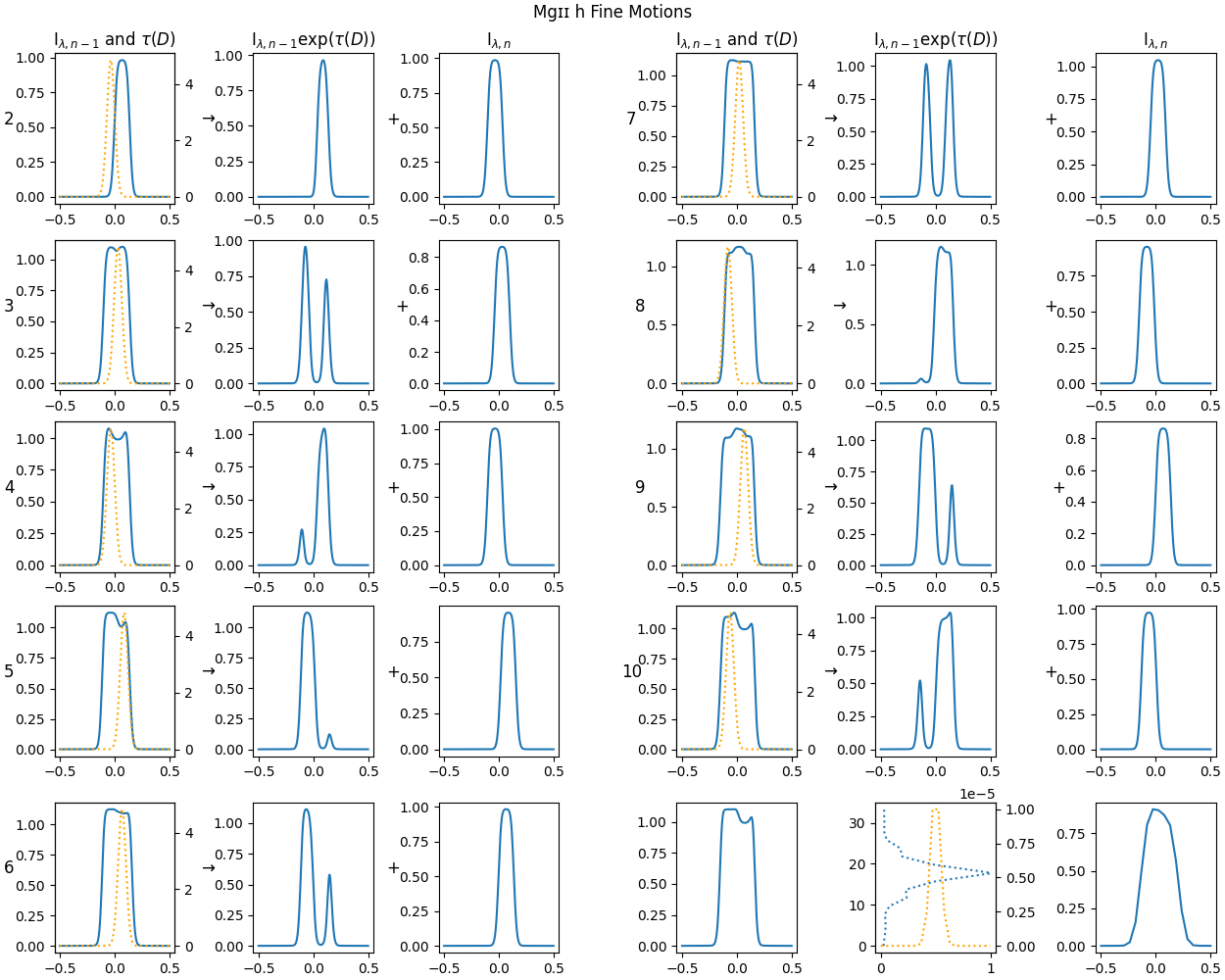}}
    \caption{Formation of line profile asymmetry. It demonstrates the summation of Eq. \ref{totalrt}. These line profiles are taken from the centre of the front-most thread. The number to the left of each set of three plots is $i$. Each set of   three plots are elements of the values in Eq. \ref{singlert} and are as follows:   $I_{i-1}$ in blue and $\tau_{i}(s)$ in dotted orange (left);   $I_{i-1}\exp\left(-\tau_{i}(s)\right)$ in blue (center);  $I_{i}$ (right). The centre and right plots are added together to create the next $I_{i-1}$. The $\lambda$ subscript denotes that these are wavelength specific intensities. The units on the y-axis are $10^5$~erg~s$^{-1}$~cm$^{-2}$~\AA$^{-1}$~sr$^{-1}$. The last three panels are different, and are as follows:   $I_{10}$, leaving the ten threads (left);  the IRIS spatial PSF, in dotted blue, where the x-axis is its value normalised such that its peak is 1 and the y-axis is parallel to the slit and the IRIS spectral PSF, in orange, where x is the normalised wavelength and y is its value normalised such that its peak is 1 (centre); the resulting line profile when convolved with the spatial and spectral PSFs of IRIS, and sampled to IRIS resolution (right). This figure originally appeared in \cite{peat2023PhD}.}
    \label{slowmgiih}
\end{figure*}

\subsection{Dynamic models}
The approach used in Sect. \ref{static} is a very simple and ideal scenario. It would be more representative of reality if the threads had random Doppler motions. \cite{gunar_lyman-line_2008} proposed that the asymmetries seen in the Lyman lines are due to {low} line-of-sight velocities. Since the optical thickness of the Lyman lines is generally much greater than \ion{Mg}{ii}~h\&k, {low} line-of-sight velocities could also be responsible for the asymmetries seen in \ion{Mg}{ii}~h\&k. More recently, \cite{tei_2020} and \cite{gunar_large_2022} suggested that line-of-sight velocities play a large role in the appearance and shape of the \ion{Mg}{ii}~h\&k lines. Following this, and using a   setup similar to that for the static models, we now add small line-of-sight velocities to investigate this effect on \ion{Mg}{ii}~h\&k. For further comparison with \cite{labrosse_radiative_2016}, we   use the same random velocities in this work. The authors drew ten velocities from a uniform distribution in the range [$-$10, 10] km~s$^{-1}$ of 7, -7, 9, -2, -7, -9, 4, -3, 4, and -7~km~s$^{-1}$ for each thread. These velocities are antiparallel {relative }to the line of sight due to the way the code defines its axes (see Fig. {\ref{setup}}).
Figure \ref{fig:spectrahv}   shows the spectra obtained from this simulation. The 1D slices show much more complex structure than in Fig. \ref{fig:spectrah}. While the exact number of threads along the line of sight is still uncertain, it is clear that there must be several.

\section{Synthetic observation}
\label{sect:obs}
Now that we have a more realistic configuration of threads, it is natural to consider how they are seen by {current} instrumentation. The only instrument currently capable of observing \ion{Mg}{ii} is the Interface Region Imaging Spectrograph \citep[IRIS;][]{depontieu_interface_2014}. IRIS has routinely been used to observe prominences in \ion{Mg}{ii}~h\&k since its launch in 2013 \citep[e.g.][]{heinzel_formation_2014, stiefel_solar_2023}; there have been many papers that attempt to infer the plasma properties from comparisons with NLTE RT codes \citep[e.g.][]{heinzel_understanding_2015, zhang_launch_2019, ruan_diagnostics_2019, tei_2020, jejcic_2022}. IRIS has also managed to decipher the motions of the previously enigmatic solar tornadoes \citep{levens_diagnostics_2018, gunar_2023}, originally observed by the EUV Imaging Spectrometer \citep[EIS;][]{culhane_euv_2007} on board Hinode \citep{kosugi_hinode_2007}. 

\begin{figure*}
    \centering
    \resizebox{\hsize}{!}
    {\includegraphics[width=\linewidth]{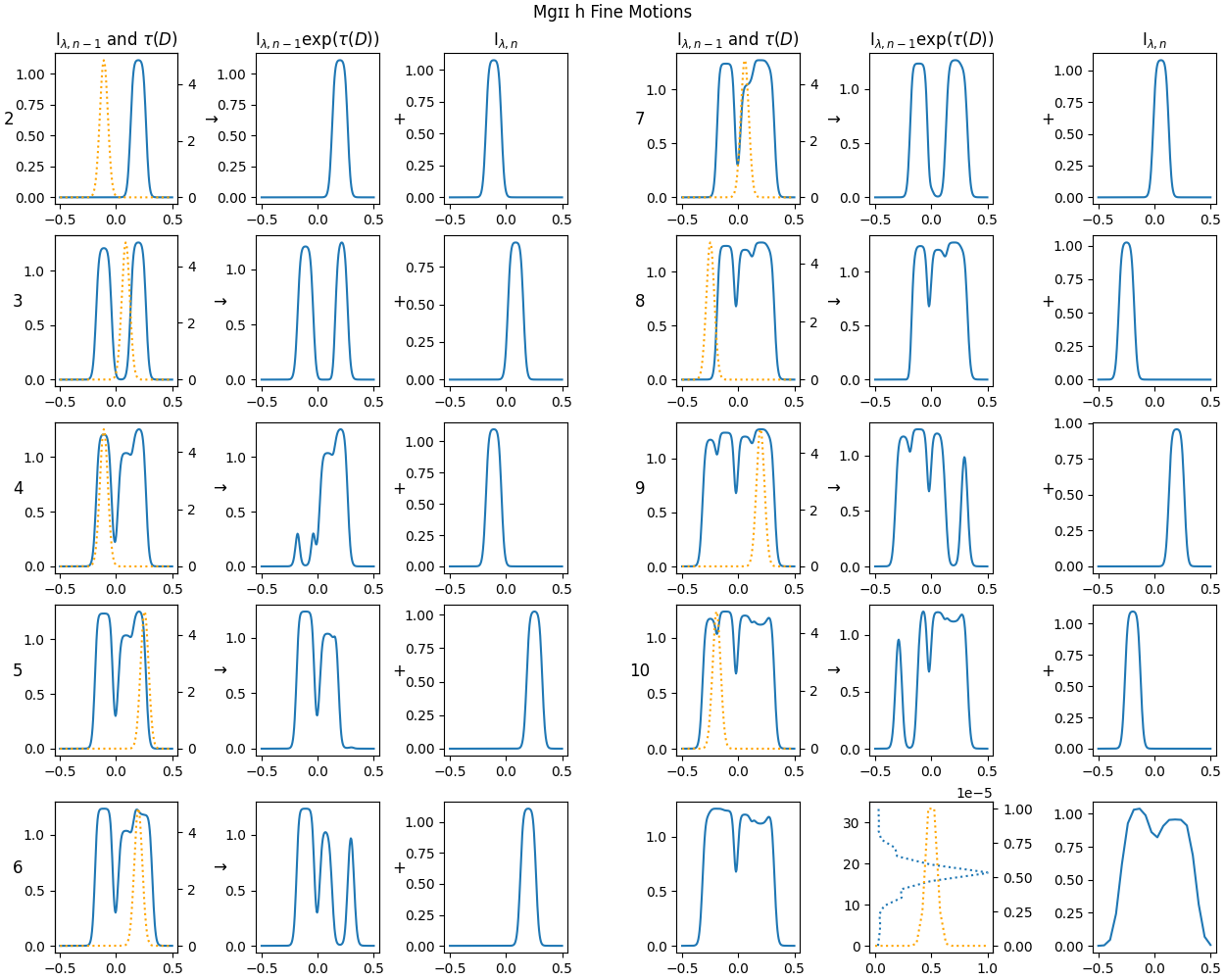}}
    \caption{Similar to Fig. \ref{slowmgiih}, but for faster moving threads. This figure originally appeared in \cite{peat2023PhD}.}
    \label{fastmgiih}
\end{figure*}

To model how IRIS would observe this fine structure, we needed to convolve our synthetic data with the point spread function(s) (PSF) of the instrument. IRIS has two PSFs to consider; the spatial PSF and the spectral PSF. The spatial PSF was determined by \cite{courrier_orbit_2018} using the transit of the planet Mercury, and is available through SolarSoft \citep[SSW;][]{freeland_data_1998}. The spatial PSF assumes a resolution of 1/6\arcsec ($\sim$$121$~km at 1AU), while our simulations had a resolution of 0.0134\arcsec ($10$~km at 1AU). Therefore, we had to resample and renormalise the spatial PSF to match the resolution of our simulations. This should also be done when deconvolving observations, but is currently ignored in the official SSW routine. This resampling was done using fourth-order weighted essentially non-oscillatory interpolation \citep[WENO4;][]{janett_novel_2019} as the shape of the PSF is better interpolated by this than other interpolation schemes. The spectral PSF is stated in \cite{depontieu_interface_2014} as being a Gaussian function with a full width at half maximum (FWHM) of two pixels. In the near-ultraviolet (NUV) filter, which observes \ion{Mg}{ii}~h\&k, this translates to a FWHM of approximately 0.1~\AA{}, or a standard deviation of 0.042~\AA{}. This may seem negligible, but it does affect the width of the line. We first convolved our simulations with the spatial PSF  as this is an artefact of the slit entrance. This is then followed by convolving with the spectral PSF  as this is due to the prism of the spectrograph. These convolutions were performed using the built-in `convolve' function of SciPy \citep{scipy}.
These PSFs can be seen in Fig. \ref{fig:psfs}.

In order to explore this open question raised by \cite{schmieder_open_2014} {and further investigate the implications of \cite{gunar_large_2022}}, we present Fig. \ref{slowmgiih}. This is similar to Fig. 4 presented in \cite{gunar_lyman-line_2008}, and very clearly demonstrates the way in which these asymmetric profiles are created. These {spectral} slices are taken from the centre of the spectra (10~000km in Fig. \ref{fig:spectrahv}). Unlike \cite{gunar_lyman-line_2008}, we do not recover striking asymmetry like that seen with{, for example,} Ly~$\beta$, {even} though the optical thickness of \ion{Mg}{ii}~h\&k is roughly two orders of magnitude lower than Ly~$\beta$. However, this is likely due to the separation of the double peaks. The separation of the double peaks of {the {principal} Lyman lines} is approximately 0.5~\AA, while most of the \ion{Mg}{ii}~h\&k line profiles produced here are single peaked. However, we do recover an interesting asymmetric line profile, but this curious detail in the shape of the profile is completely obfuscated by the PSFs of the instrument, ultimately resembling a standard single-peaked \ion{Mg}{ii}~h profile. This demonstrates that the PSF needs to be carefully considered when analysing IRIS data.

\begin{figure}
    \centering
    \includegraphics[width=\linewidth]{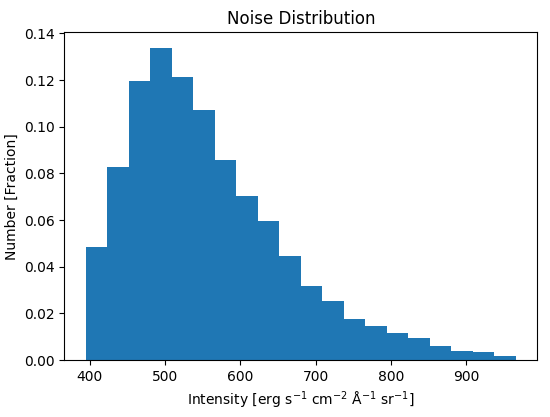}
    \caption{Histogram of the mean intensities between 2800.33~\AA{} and 2801.91~\AA{}.}
    \label{fig:noisehist}
\end{figure}

\begin{figure}
    \centering
    \includegraphics[width=\linewidth]{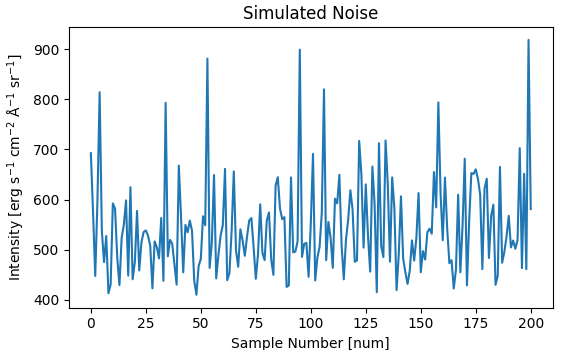}
    \caption{Samples from the distribution in Fig. \ref{fig:noisehist} (201 in total). This represents our synthetic noise.}
    \label{fig:fakenoise}
\end{figure}

{However, more complex and/or double-peaked profiles are routinely observed in IRIS prominence spectra \citep[e.g.][]{levens_structure_2016, jejcic_statistical_2018}. Additionally, prominences routinely display line-of-sight velocities of up to 30~km~s$^{-1}$ \citep{labrosse_physics_2010}; therefore, it could be argued that an increase in the implemented line-of-sight velocities may produce the more striking asymmetries discussed in \cite{schmieder_open_2014}}.
To attempt to produce {these} more striking asymmetries, we implemented greater line-of-sight velocities. We increased the velocities by a factor of three, resulting in 21~km~s$^{-1}$, -21~km~s$^{-1}$, 27~km~s$^{-1}$, -6~km~s$^{-1}$, -21~km~s$^{-1}$, 27~km~s$^{-1}$, 12~km~s$^{-1}$, -9~km~s$^{-1}$, 12~km~s$^{-1}$, and -21 ~km~s$^{-1}$. From this we obtained a fascinating dip in the middle of the line profile (see Fig. \ref{fastmgiih}). This is due to the larger range of velocities where approximately half of the threads do not meaningfully interact with one another. This effect can clearly be seen in Fig. \ref{fastmgiih}. However, once again, the PSFs of the instrument smooths out this interesting feature. Unlike the slower case, here it produces a much more asymmetrical line profile similar to that commonly observed by IRIS. This confirms that one mechanism to produce asymmetrical \ion{Mg}{ii}~h\&k line profiles is the presence of fine structure with independent large velocities. 
{While these profiles were convolved with the PSFs of IRIS, no noise was included. If we wanted to use this as if it were a real observation in Sect. \ref{modelling}, noise would have to be considered. Noise would have a non-negligible effect on the result of the deconvolution.} To do this, we used the \ion{Mg}{ii}~h\&k IRIS observations of the prominence of {19 April 2018} (presented in \citeauthor{peat_solar_2021} \citeyear{peat_solar_2021}, \citeauthor{barczynski_spectro-imagery_2021} \citeyear{barczynski_spectro-imagery_2021}, and \citeauthor{labrosse_first_2022} \citeyear{labrosse_first_2022}). Using the first raster, we took the mean of the signal between 2800.33~\AA{} and 2801.91~\AA{} of every respective pixel. This part of the spectrum is said to be photospheric, and so no emission should be present here when observing off-limb. This produced approximately 6240 measurements of the noise. To remove outliers, any noise level lower than the 3rd percentile and higher than the 97th percentile were removed. From this, we plotted a normalised histogram (see Fig. \ref{fig:noisehist}) from which we drew samples to create synthetic noise (see    Fig. \ref{fig:fakenoise}).
\begin{table}
\centering
\begin{tabular}{ccc} \hline\hline
Parameter        & Unit           & Value                                                                                                                                                      \\
\hline
$T_{\text{cen}}$ & kK              & \begin{tabular}[c]{@{}c@{}}6, 8, 10, 12, 15\\ 20, 25, 35, 40\end{tabular} \\
$T_{\text{tr}}$  & kK              & 100                                                                                                                                              \\
$p_{\text{cen}}$ & dyne cm$^{-2}$ & \begin{tabular}[c]{@{}c@{}}0.01, 0.02, 0.05\\ 0.1, 0.2, 0.5, 1 \end{tabular}\\
                                                                                                                       
$p_{\text{tr}}$  & dyne cm$^{-2}$ & 0.01                                                                                                                                                       \\
Slab Width       & km             & 45 -- 124~100                                                                                                                                         \\
M                & g cm$^{-2}$    & 3.7$\times10^{-8}$ -- 5.1$\times10^{-4}$                                                                                                                   \\
H                & Mm             & 10, 30, 50                                                                                                                                        \\
$v_{\text{T}}$   & km s$^{-1}$    & 5, 8, 13                                                                                                                                                   \\
$v_{\text{rad}}$ & km s$^{-1}$    & \begin{tabular}[c]{@{}c@{}}0, 2, 4, 6, 8, 10, 20\\40, 60, 80, 100, 150, 200 \end{tabular}\\

$\gamma$         &                & 0, 2, 4, 5, 10        \\ \hline \hline                                                                                                                    
\end{tabular}
\caption{Model Parameters. A value of $\gamma$ is only valid for PCTR models {and $\gamma$=0 is used to identify an isothermal and isobaric model; $\gamma$=0 is mathematically meaningless}. Note that not every combination of these parameters is present in the grid of models.}
\label{params1d}
\end{table}
{Before adding noise, the spectral window of the synthetic observations were} increased to 3~\AA{} to better simulate what is done when using xRMS on real observations. This was simply done by padding the simulation with zeroes. The synthetic noise was {then} added to the data by drawing random samples from the distribution in Fig. \ref{fig:noisehist}. {The middle plots of Fig. \ref{fig:deconh} show the final modelled line profiles from Figs. \ref{slowmgiih} and \ref{fastmgiih} convolved with the PSFs, and sampled to the spectral resolution  of IRIS with noise. An interesting observation  is that the peak intensities drop by approximately 20\% after the profiles are reduced to the resolution of IRIS. However, this is to be expected, as the resampling process involves the calculation of a mean that   causes the peak intensity of functions of this shape to fall. From here on  we   treat these line profiles as if they are our observations and attempt to invert them using 1D forward modelling.}

\section{One-dimensional forward modelling and inversion}
\label{modelling}

These simulations carry some consequences for using 1D modelling to attempt to recover the thermodynamic properties of prominences. To demonstrate this, we   treat these simulated line profiles as if they were IRIS observations and use the xRMS method \citep[][]{peatinprep}, an improved version of the rolling root mean square  \citep[rRMS;][]{peat_solar_2021} method, to attempt to recover the thermodynamic properties of these simulated observations.  


\begin{figure*}
    \centering
    \includegraphics[width=\linewidth]{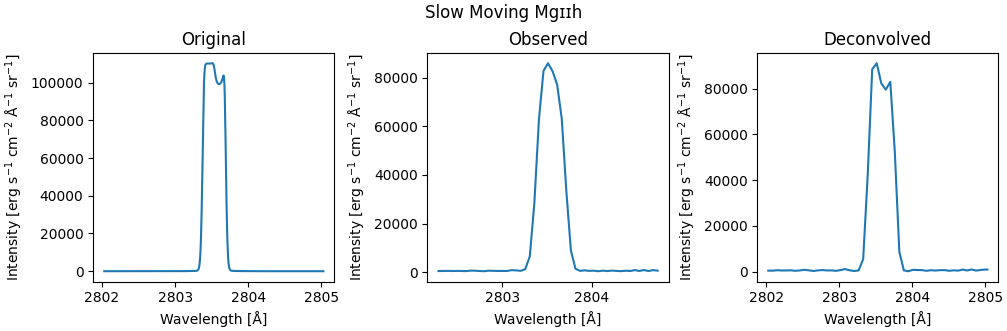}
    \includegraphics[width=\linewidth]{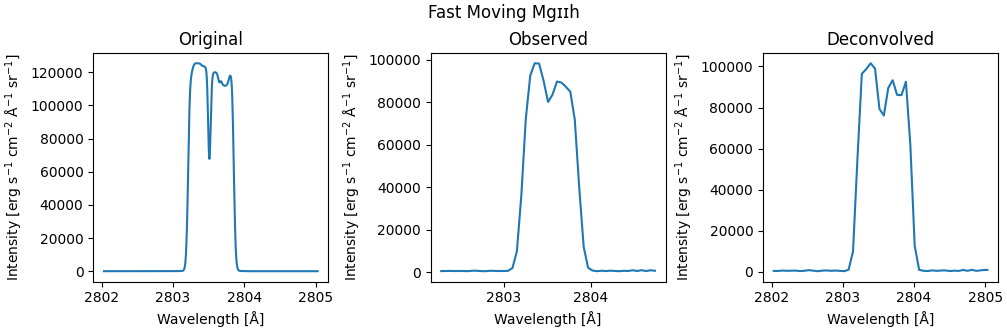}
    \caption{Low- and high-velocity synthetic observations. The upper panels show the low-velocity simulations and the lower panels show the high-velocity simulations. The left panels show the original line profiles. The middle panels show the line profiles as they would be observed by IRIS. The right panels show the line profiles after {the Richardson--Lucy deconvolution}.} 
    \label{fig:deconh}
\end{figure*}

We used a grid of 23~940 1D {slab} NLTE \ion{Mg}{ii}~h\&k models from PROM \citep{gouttebroze_hydrogen_1993, heinzel_theoretical_1994, levens_modelling_2019}.
The parameters of these models can be seen in Table \ref{params1d}. These models include a mixture of isothermal and isobaric atmospheres, and those containing a prominence-to-corona transition region (PCTR). The parameters $T_{\text{cen}}$ and $p_{\text{cen}}$ are the central temperature and pressure, respectively; $T_{\text{tr}}$ and $p_{\text{tr}}$ are the temperature and pressure at the edge of the PCTR, respectively; slab width is the width of the slab; $M$ is the column mass; $H$ is the height above the solar surface; $v_T$ is the microturbulent velocity; $v_{\text{rad}}$ is the outward radial velocity of the slab {(i.e. perpendicular to the solar surface)}; and $\gamma$ is a dimensionless number that dictates the extent of the PCTR. A $\gamma$ value of 0 indicates the model is isothermal and isobaric {(i.e. Eqs. \ref{tstrat} and \ref{pstrat} do not apply). In these} isothermal and isobaric models, {$T=T_{\text{cen}}$ and $p=p_{\text{cen}}$}. For non-zero values of $\gamma$, the lower the value of $\gamma$, the more extended the PCTR is. The PCTR here is formulated as a function of column mass, as in \cite{anzer_energy_1999},
\begin{equation} 
    T(m)=T_{\text{cen}}+(T_{\text{tr}}-T_{\text{cen}})\left(1-4\frac{m}{M}\left(1-\frac{m}{M}\right)\right)^\gamma,
    \label{tstrat}
\end{equation}
\begin{equation}
    p(m)=4p_c\frac{m}{M}\left(1-\frac{m}{M}\right)+p_{\text{tr}},
    \label{pstrat}
\end{equation}
for $\gamma\geq2$, and where $p_c=p_{\text{cen}}-p_{\text{tr}}$. 

With observations, the first step when preparing the data for use with xRMS is radiometric calibration. However, the simulated observations are already in units of intensity and do not need to be calibrated.

After radiometric calibration, the data is deconvolved from the PSFs of the instrument. This is achieved through the use of a Richardson--Lucy deconvolution with ten iterations \citep{richardson1972, lucy1974}. This {is a robust} deconvolution method {that is} superior to other deconvolution schemes in the presence of noise \citep{fish1995}. The implementation of this Richardson--Lucy deconvolution is {identical to} the official SSW routine for the deconvolution of the spatial PSF. However, before the deconvolution was performed, we resampled the spatial PSF such that it had the same resolution as our synthetic observations (see  Sect. \ref{sect:obs}). This was carried out on both the low- and high-velocity simulations.  The results of these operations can be seen in Fig. \ref{fig:deconh}. The interesting features, which become smoothed out due to the PSFs of the instrument, can be recovered to a satisfactory extent by the Richardson--Lucy deconvolution. The loss of detail here appears to be a consequence of the resolution of the instrument and not of the deconvolution. 

Our synthetic observations were then supplied to xRMS as if they were data prepared from real observations. xRMS works by taking the cross-correlation of the models and the data. When attempting to minimise a mean square, a cross-correlation is naturally produced \citep{elliott_handbook_1987}, and therefore is also produced when minimising a root mean square. Using a cross-correlation offers a significant increase in computation speed over the original method described in \cite{peat_solar_2021}. 
When attempting to find the best match between line profiles and the models, xRMS selects the model that produces the {lowest sum of the RMS between the \ion{Mg}{ii}~h\&k synthetic profiles and the observed \ion{Mg}{ii}~h\&k profiles.} For a model to be classified as `satisfactory' the RMS sum must be lower than 15~000 \citep{peat_solar_2021}.

\begin{figure}
    \centering
    \includegraphics[width=\linewidth]{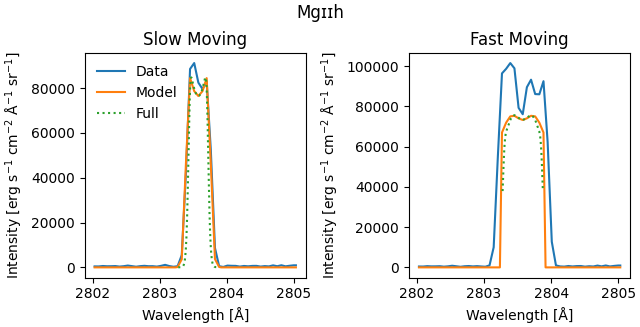}
    \caption{One-dimensional  models found by xRMS that return the lowest RMS. {The green dotted lines labelled `Full' are the model profiles before downsampling and zero-padding. Both of these models are isothermal and isobaric.} \textit{Left:} Slow moving models. The parameters of the fit are $T$=8~kK; $p$=0.1~dyn~cm$^{-2}$; slab width=1000~km; H=10~Mm; $v_\text{T}$=8~km~s$^{-1}$; and $v_\text{rad}$=8~km~s$^{-1}$. \textit{Right:} Fast moving models. The parameters of the fit are $T$=6~kK;  $p$=0.5~dyn~cm$^{-2}$; slab width=2000~km; H=10~Mm; $v_\text{T}$=13~km~s$^{-1}$; and $v_\text{rad}$=0~km~s$^{-1}$.}
    \label{fig:hmatches}
\end{figure}

For the low-velocity models, the best fit model had an RMS of 8846.72. For the high-velocity models, the best fit model had an RMS of 38435.93,  which was classified as unsatisfactory. These results can be seen in Figure \ref{fig:hmatches}. Both of these best fitting models are found to be isothermal and isobaric. This is understandable as the drop in density in the PCTR causes very little to no \ion{Mg}{ii}~h\&k emission in the PCTR. This implies that the emission seen is no different to that of an isothermal and isobaric atmosphere. Additionally, this type of model is typically found as the best match for regions containing complex line profiles \citep{peat_solar_2021}. 
The increase in microturbulent velocity found for the faster models is due to the large line width. It was argued in \cite{peat_solar_2021} that large microturbulent velocities could be used in 1D forward modelling to attempt to account for unresolved fine motions that contribute to an increased line width. However, as we can see in Fig. \ref{fig:hmatches}, while the line width has been accounted for, the line itself is far too complex for 1D models to sufficiently reproduce. 
Meanwhile, for the slower models, the match found is very reasonable. It correctly recovers the pressure and approximate height of the individual threads. However, the other values are incorrect. The simulated prominence has a core temperature of 6~kK and the PCTR is 100~kK; xRMS finds an isothermal temperature of 8~kK. 
However, we could instead be running into the issue of degeneracy. It can be shown that 1D single-species forward modelling produces degenerate solutions for the inferred thermodynamic properties {\citep{ruan_diagnostics_2019, jejcic_2022}.   \cite{jejcic_2022} demonstrate that the use of multispecies diagnostics can  remove these degeneracies when using 1D models. Therefore, future studies should endeavour to use multiple species and attempt to match the line profiles point-for-point. Additionally, with recent advances in computing, a Bayesian approach using methods such as diffusive nested sampling \citep[DNest;][]{brewer2011diffusive} may now be practical.}

\section{Conclusions}
The integrated intensity along the slit of a spectrograph may give information pertaining to the multithread nature of the plasma that produced the spectra. This information is more readily recovered than just by the slit spectra alone. 

Asymmetrical line profiles observed by IRIS may be a consequence of the fine unresolved motions within the plasma. As such, careful consideration must be taken when attempting to recover the plasma diagnostics of these lines when using 1D models. Even though good matches can be found for some of these spectra, it does not necessarily mean that the correct diagnostics have been recovered. Due to a combination of the PSF and spectral resolution of the instrument, some of the peculiar structure and detail is lost. {No   algorithm will be able to recover this lost information.} This can further compound the above-mentioned issue of the use of 1D models to recover the plasma diagnostics.

In future attempts to recover the plasma properties, the structure of the plasma should be carefully considered such that erroneous diagnostics are not recovered through 1D modelling. 
Using 1D models, it is difficult to obtain diagnostics for sections of the prominence where the line-of-sight intercepts large and/or dense sections of the prominence. The asymmetrical profiles seen in these areas, which cannot be accurately represented by 1D models, may be due to independently moving fine structures. {When using 1D models, it is crucial to use multiple species from coordinated observation in order to better constrain the plasma diagnostics \citep{jejcic_2022}.}

Additionally, many new models of higher dimensionality are being created \citep[such as][]{gunar_3d_2015,jenkins_non_2023} that could be used in future when attempting to recover the plasma parameters. {Perhaps in conjunction with a Bayesian approach such as  diffusive nested sampling \citep[DNest;][]{brewer2011diffusive}.}

\begin{acknowledgements}
      AWP acknowledges financial support from STFC via grant ST/S505390/1. NL acknowledges support from STFC grant ST/T000422/1. We would also like to thank Dr Christopher M. J. Osborne for writing the Python implementation of WENO4 (https://github.com/Goobley/Weno4Interpolation) and Dr Graham S. Kerr for supplying the IRIS spatial PSFs.
      IRIS is a NASA small explorer mission developed and operated by LMSAL with mission operations executed at NASA Ames Research Center and major contributions to downlink communications funded by ESA and the Norwegian Space Centre.
      This research used version 3.5.2 of Matplotlib \citep{matplotlib}, version 1.22.4 of NumPy \citep{harris_array_2020}, version 1.8.1 of SciPy \citep{scipy}, and version 5.1 of Astropy, (http://www.astropy.org) a community-developed core Python package for Astronomy \citep{the_astropy_collaboration_astropy_2013, the_astropy_collaboration_astropy_2018}.
\end{acknowledgements}

\bibliography{thesis}
\bibliographystyle{aa}

\newpage
\appendix

\section{Supplementary \ion{Mg}{ii}~k plots}
\begin{figure}[h]
    \centering
    \includegraphics[width=\linewidth]{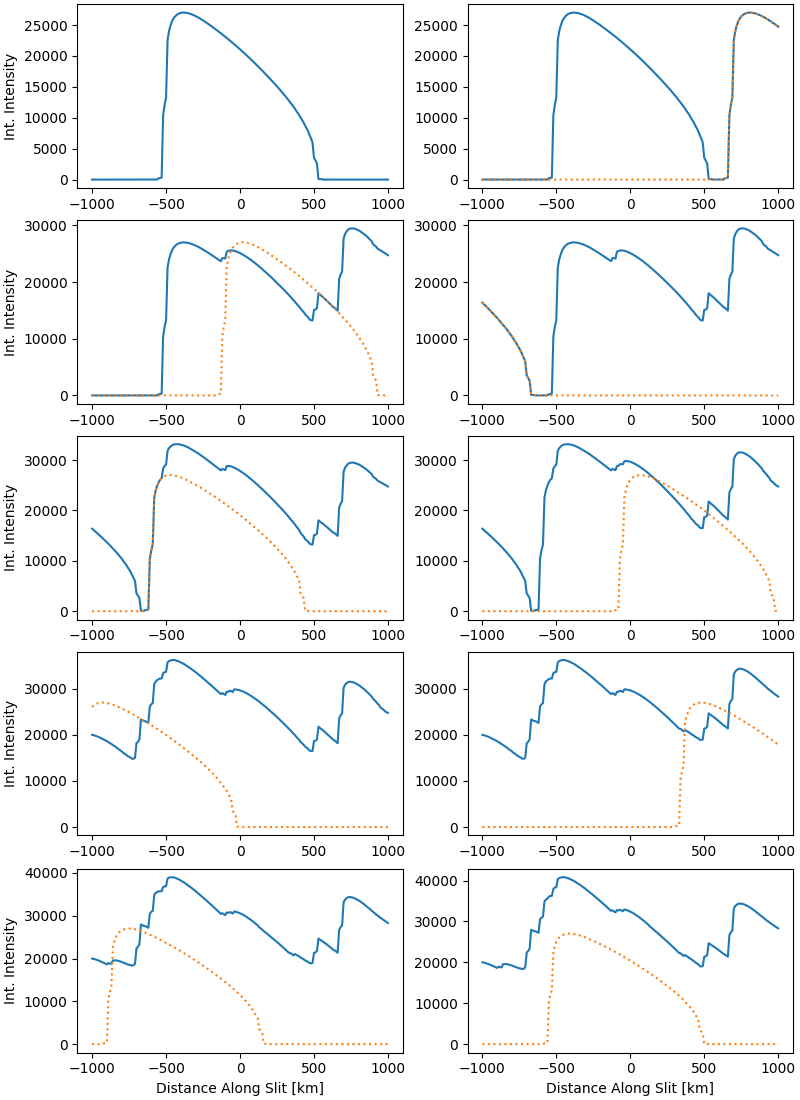}
    \caption{Same as Fig. \ref{fig:slith}, but for \ion{Mg}{ii}~k. The units on the y-axis are erg~s$^{-1}$~cm$^{-2}$~sr$^{-1}$. This plot originally appeared in \cite{peat2023PhD}.}
    \label{fig:slitk}
\end{figure}
\vfill\break
\begin{figure}[h]
    \centering
    \includegraphics[width=\linewidth]{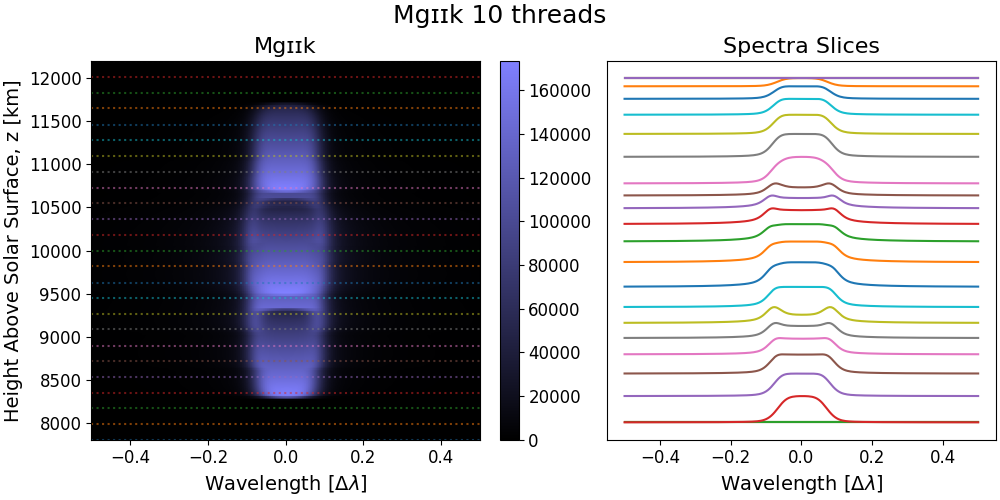}
    \caption{Same as Fig. \ref{fig:spectrah}, but  for \ion{Mg}{ii}~k. The units of the colour bar are erg~s$^{-1}$~cm$^{-2}$~\AA$^{-1}$~sr$^{-1}$. This figure originally appeared in \cite{peat2023PhD}.}
    \label{fig:spectrak}
\end{figure}

\begin{figure}[h]
    \centering
    \includegraphics[width=\linewidth]{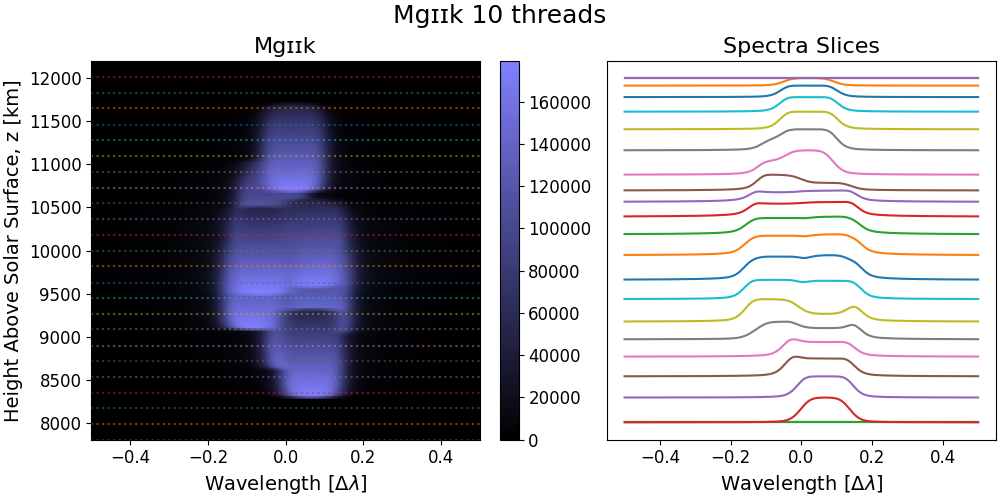}
    \caption{Same as Fig. \ref{fig:spectrahv}, but  for \ion{Mg}{ii}~k. The units of the colour bar are erg~s$^{-1}$~cm$^{-2}$~\AA$^{-1}$~sr$^{-1}$.}
    \label{fig:spectrakv}
\end{figure}

\newpage

\begin{figure*}[h]
    \centering
    \resizebox{\hsize}{!}
    {\includegraphics[width=\linewidth]{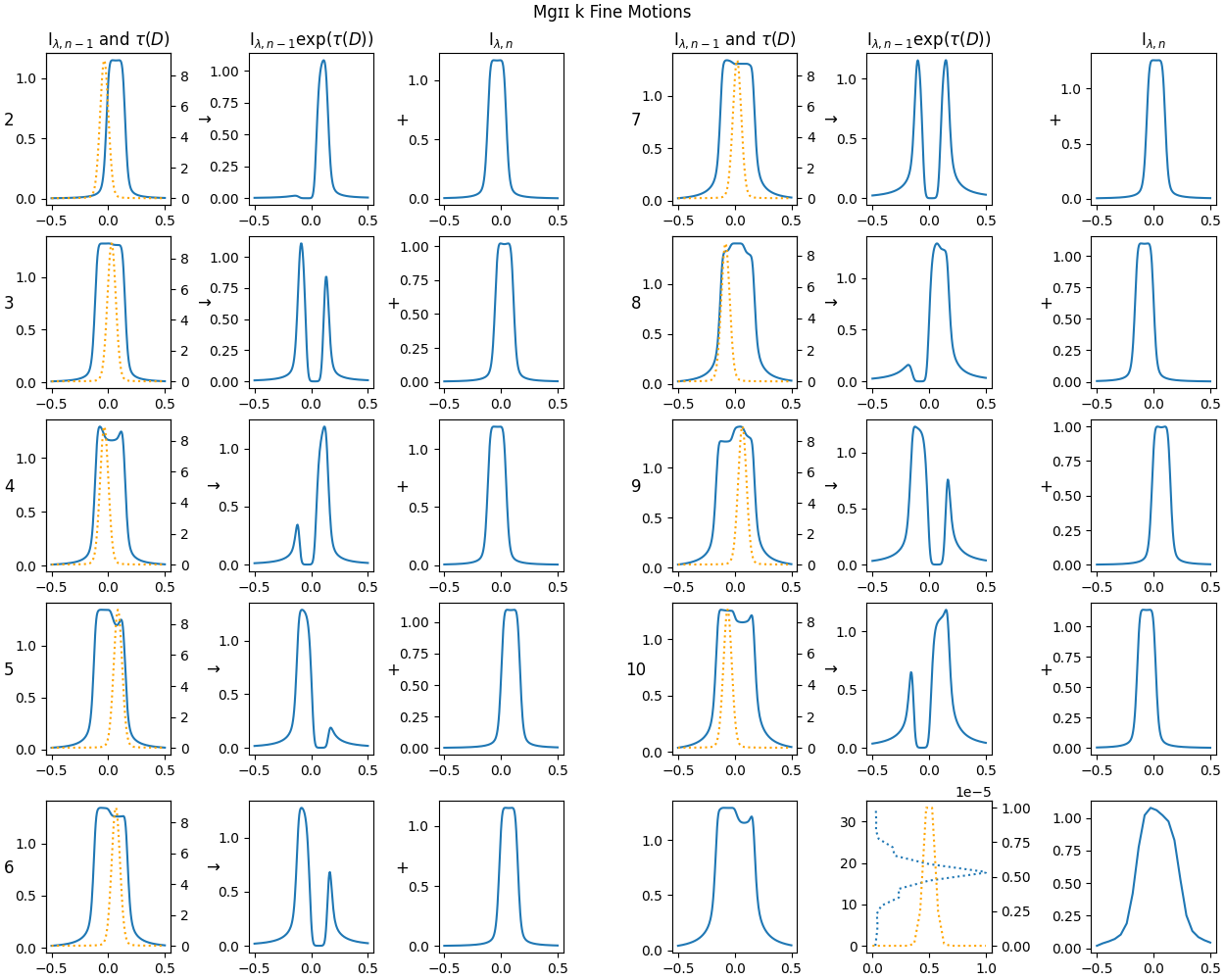}}
    \caption{Same as Fig. \ref{slowmgiih}, but  for \ion{Mg}{ii}~k. The units on the y-axes are $10^5$~erg~s$^{-1}$~cm$^{-2}$~\AA$^{-1}$~sr$^{-1}$. This figure originally appeared in \cite{peat2023PhD}.} 
    \label{slowmgiik}
\end{figure*}

\begin{figure*}
    \centering
    \resizebox{\hsize}{!}
    {\includegraphics[width=\linewidth]{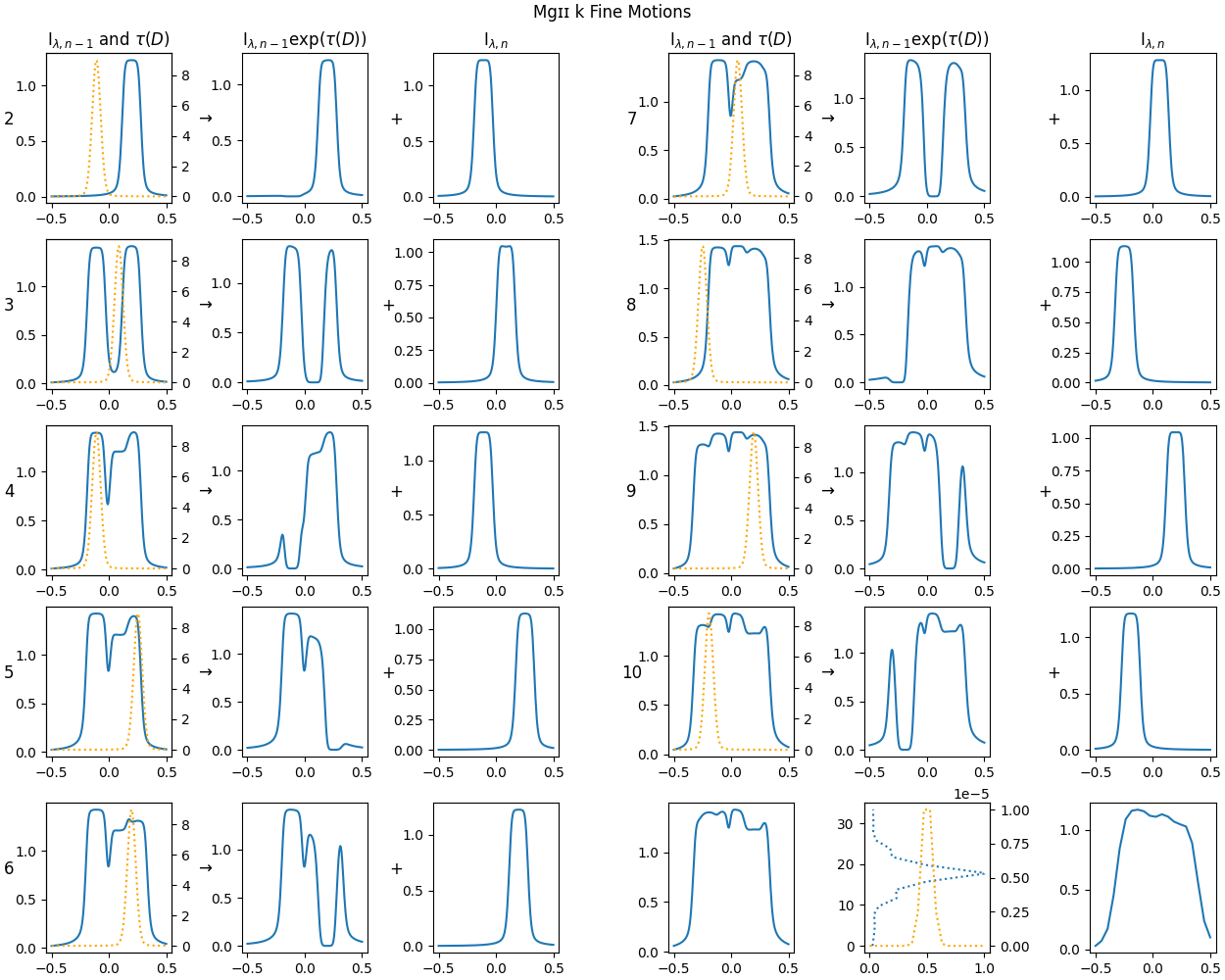}}
    \caption{Same as Fig. \ref{fastmgiih}, but  for \ion{Mg}{ii}~k. The units on the y-axes are $10^5$~erg~s$^{-1}$~cm$^{-2}$~\AA$^{-1}$~sr$^{-1}$. This figure originally appeared in \cite{peat2023PhD}.}
    \label{fastmgiik}
\end{figure*}

\begin{figure*}
    \centering
    \includegraphics[width=\linewidth]{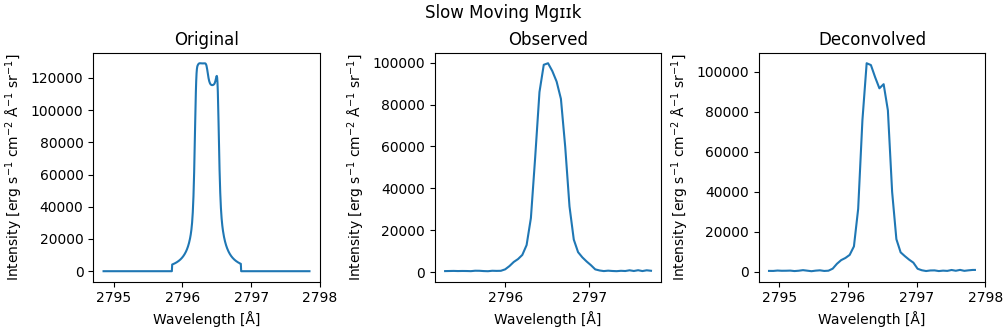}
    \includegraphics[width=\linewidth]{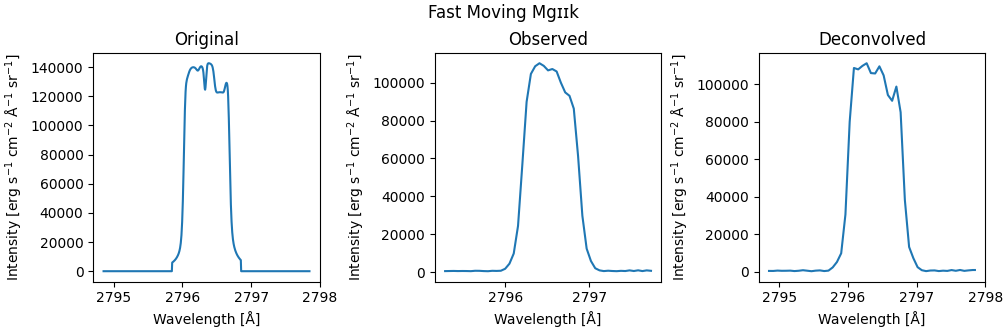}
    \caption{Same as Fig. \ref{fig:deconh}, but  for \ion{Mg}{ii}~k.}
    \label{fig:deconk}

    \centering
    \includegraphics[width=0.7\linewidth]{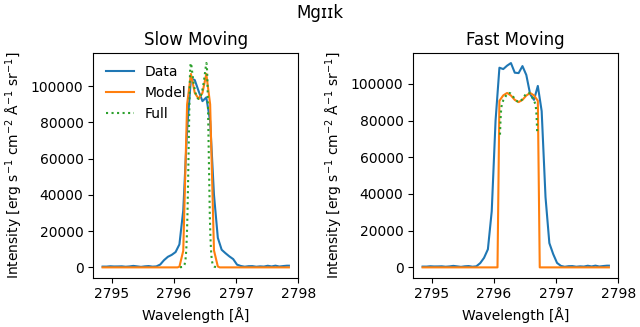}
    \caption{Same as Fig. \ref{fig:hmatches}, but  for \ion{Mg}{ii}~k.}
    \label{fig:kmatches}
\end{figure*}
 
\end{document}